# Regge behaviour of distribution functions and evolution of gluon distribution function in Next-to-Leading order at low-x

U. Jamil[1] and J. K. Sarma[2]

Department of Physics, Tezpur University, Napam, Tezpur-784028, Assam, India

**Abstract.** Evolution of gluon distribution function from Dokshitzer-Gribov-Lipatov-Altarelli-Parisi (DGLAP) evolution equation in next-to-leading order (NLO) at low-x is presented assuming the Regge behaviour of quarks and gluons at this limit. We compare our results of gluon distribution function with MRST2004, GRV98LO and GRV98NLO parameterizations and show the compatibility of Regge behaviour of quark and gluon distribution functions with perturbative quantum chromodynamics (PQCD) at low-x.

## 1 Introduction

In our earlier communication [1], we derived the solution of DGLAP evolution equation for gluon distribution function in leading order (LO) at low-x considering Regge behaviour of distribution functions. Here, in continuation of the earlier work we solved DGLAP evolution equation for gluon distribution function at low-x in next-to-leading order (NLO) and the t and x-evolutions of gluon distribution function thus obtained have been compared with global MRST2004 and GRV98 parameterizations. In PQCD, since the higher order terms in the leading logarithmic series $(\alpha_S(Q^2)\ln Q^2)^n$ are important, so, simply adding the leading order contribution of the quark and gluon is not enough to get the parton distributions. But at large $Q^2$ ($Q^2 >> \Lambda^2$), which is the requirement of DGLAP equation, running coupling constant $\alpha_S(Q^2)$ is small and contribution of higher order terms decrease very fast. At our current scale of $Q^2$ though $Q^2 >> \Lambda^2$, $\alpha_S(Q^2)$ is not that much small. So higher order terms, at least up to NLO, are also important [2-4]. Here, section 1, section 2, section 3 and section 4 are the introduction, theory, results and discussion, and conclusions respectively.

---

[1]jamil@tezu.ernet.in, [2]jks@tezu.ernet.in

## 2 Theory

The DGLAP evolution equation for gluon distribution function upto NLO has the standard form [5, 6]

$$Q^2 \frac{\partial}{\partial Q^2} G(x, Q^2) = \frac{\alpha_s(Q^2)}{2\pi} \int_x^1 \left[ P_{gg}^1(\omega) G(x/\omega, Q^2) + P_{gq}^1(\omega) F_2^S(x/\omega, Q^2) \right] d\omega$$

$$+ \left[ \frac{\alpha_s(Q^2)}{2\pi} \right]^2 \int_x^1 \left[ P_{gg}^2(\omega) G(x/\omega, Q^2) + P_{gq}^2(\omega) F_2^S(x/\omega, Q^2) \right] d\omega \quad (1)$$

where $\alpha_s(Q^2) = \frac{4\pi}{\beta_0 Q^2} \left[ 1 - \frac{\beta_1 \ln Q^2}{\beta_0^2 Q^2} \right]$, $N_f$ being the number of flavours and $\Lambda$ is the QCD cut-off parameter depends on the renormalization scheme, $\hat{a}_0$ and $\hat{a}_1$ are the expansion coefficient of the $\hat{a}$-function and they are given by $\beta_0 = \frac{33 - 2n_f}{3}$, $\beta_1 = \frac{306 - 38n_f}{3}$. The splitting functions have their forms as given in [5, 6]. At low-x, for x → 0, only gluon splitting function matters [6, 7]. So keeping the full form of other splitting functions, we make some approximation of the splitting function $P^2_{gg}(\omega)$, retaining only its leading term as x → 0 [6], i.e., we take $P_{gg}^2(\omega) \cong \frac{52}{3} \frac{1}{\omega}$. After changing the variable $Q^2$ by t, where $t = \ln(Q^2/\Lambda^2)$ and putting the respective kernels we get

$$\frac{\partial G(x, t)}{\partial t} = \frac{\alpha_S(t)}{2\pi} \left\{ 6 \times \left( \frac{11}{12} - \frac{N_f}{18} + \ln(1-x) \right) G(x, t) + 6 \times I_g \right\}$$

$$+ \left[ \frac{\alpha_s(Q^2)}{2\pi} \right]^2 \int_x^1 \left[ \frac{52}{3} \frac{1}{\omega} G(x/\omega, Q^2) + A(\omega) F_2^S(x/\omega, Q^2) \right] d\omega, \quad (2)$$

where

$$I_g = \int_x^1 d\omega \left[ \frac{\omega G(x/\omega, t) - G(x,t)}{1-\omega} + \left( \omega(1-\omega) + \frac{1-\omega}{\omega} \right) G(x/\omega, t) + \frac{2}{9} \left( \frac{1+(1-\omega)^2}{\omega} \right) F_2^S(x/\omega, t) \right],$$

and

$$A(\omega) = C_F^2 . A_1(\omega) + C_F . C_G . A_2(\omega) + C_F . T_R . N_F . A_3(\omega),$$

$$A_1(\omega) = -\frac{5}{2} - \frac{7}{2}\omega + \left(2 + \frac{7}{2}\omega\right)\ln\omega + \left(-1 + \frac{1}{2}\omega\right)\ln^2\omega - 2\omega\ln(1-\omega)$$

$$+ \left(-3\ln(1-\omega) - \ln^2(1-\omega)\right)\frac{1+(1-\omega)^2}{\omega},$$

$$A_2(\omega) = \frac{28}{9} + \frac{65}{18}\omega + \frac{44}{9}\omega^2 + \left(-12 - 5\omega - \frac{8}{3}\omega^2\right)\ln\omega + (4+\omega)\ln^2\omega + 2\omega\ln(1-\omega)$$

$$+ \left(-2\ln\omega\ln(1-\omega) + \frac{1}{2}\ln^2\omega + \frac{11}{3}\ln(1-\omega) + \ln^2(1-\omega) - \frac{1}{6}\pi^2 + \frac{1}{2}\right)\frac{1+(1-\omega)^2}{\omega}$$

$$- \frac{1+(1+\omega)^2}{\omega}\int_{\omega/1-\omega}^{1/1-\omega}\frac{dz}{z}\ln\frac{1-z}{z}$$

and

$$A_3(\omega) = -\frac{4}{3}\omega - \left(\frac{20}{9} + \frac{4}{3}\ln(1-\omega)\right)\frac{1+(1-\omega)^2}{\omega}.$$

$C_A$, $C_G$, $C_F$, and $T_R$ are constants associated with the color SU (3) group and $C_A = C_G = N_C = 3$, $C_F = (N_C^2-1)/2N_C$ and $T_R = 1/2$. $N_C$ is the number of colours.

Now let us consider the Regge behaviour of gluon distribution function [1, 8-10] as

$$G(x, t) = T(t)x^{-\lambda} \qquad (3)$$

Since the DGLAP evolution equations of gluon and singlet structure functions in leading and next-to-leading order are in the same forms of derivative with respect to t, so we consider the ansatz [1] for simplicity,

$$G(x, t) = K(x)F_2^s(x, t) \qquad (4)$$

where K(x) is a parameter to be determined from phenomenological analysis and we assume $K(x) = K$, $ax^b$ or $ce^{dx}$ where K, a, b, c and d are constants. Now,

$$F_2^S(x/\omega, t) = \frac{1}{K(x/\omega)}G(x/\omega, t) = \frac{\omega^\lambda}{K(x/\omega)}G(x, t). \qquad (5)$$

Putting equations (3) and (5) in equation (2), we get

$$\frac{\partial G(x,t)}{\partial t} - G(x,t).P(x,t) = 0, \qquad (6)$$

where

$$P(x,t) = \frac{\alpha_s(t)}{2\pi} \cdot f_1(x) + \left(\frac{\alpha_s(t)}{2\pi}\right)^2 \cdot f_2(x),$$

$$f_1(x) = 6 \times \left(\frac{11}{12} - \frac{N_f}{18} + \ln(1-x)\right) + 6 \times \int_x^1 d\omega \left[\frac{(\omega^{\lambda+1} - 1)}{1-\omega} + \left(\omega(1-\omega) + \frac{1-\omega}{\omega}\right)\omega^\lambda + \frac{2}{9}\left(\frac{1+(1-\omega)^2}{\omega}\right)\frac{\omega^\lambda}{K(x/\omega)}\right]$$

and

$$f_2(x) = \int_x^1 \left[\frac{52}{3}\omega^{\lambda-1} + A(\omega)\frac{\omega^\lambda}{K(x/\omega)}\right] d\omega.$$

For possible solutions in NLO, we have taken the expression for $\left(\frac{\alpha_s(t)}{2\pi}\right)$ upto LO correction and we have to put an extra assumption $\left(\frac{\alpha_s(t)}{2\pi}\right)^2 = T_0\left(\frac{\alpha_s(t)}{2\pi}\right)$ [11, 12], where $T_0$ is a numerical parameter. But $T_0$ is not arbitrary. We choose $T_0$ such that difference between $T^2(t)$ and $T_0 T(t)$ is minimum in the region of our discussion. ( fig.1(a) ). Equation (6) reduces to

$$\frac{\partial G(x,t)}{\partial t} - \frac{G(x,t)}{t} \cdot P(x) = 0 \tag{7}$$

with

$$P(x) = \frac{2}{\beta_0} \cdot f_1(x) + T_0 \cdot \frac{2}{\beta_0} \cdot f_2(x)$$

Integrating equation (7) we get

$$G(x, t) = C \cdot t^{P(x)}, \tag{8}$$

where C is a constant of integration.

Applying initial conditions at $x = x_0$, $G(x, t) = G(x_0, t)$, and at $t = t_0$, $G(x, t) = G(x, t_0)$, we found the t and x-evolution equations for the gluon distribution function in NLO respectively as

$$G(x, t) = G(x, t_0)(t/t_0)^{P(x)} \tag{9}$$

and

$$G(x, t) = G(x_0, t) t^{\{P(x) - P(x_0)\}}. \tag{10}$$

Now ignoring the quark contribution to the gluon distribution function we get from the evolution equation (2)

$$\frac{\partial G(x,t)}{\partial t} = \frac{\alpha_S(t)}{2\pi}\left\{6\times\left(\frac{11}{12} - \frac{N_f}{18} + \ln(1-x)\right)G(x,t) + 6\times I'_g\right\}$$

$$+ \left[\frac{\alpha_s(Q^2)}{2\pi}\right]^2 \int_x^1 \left[\frac{52}{3}\frac{1}{\omega}G(x/\omega, Q^2)\right]d\omega, \qquad (11)$$

where

$$I'_g = \int_x^1 d\omega\left[\frac{\omega G(x/\omega,t) - G(x,t)}{1-\omega} + \left(\omega(1-\omega) + \frac{1-\omega}{\omega}\right)G(x/\omega,t)\right].$$

pursuing the same procedure as above, we get the t and x-evolution equations for the gluon distribution function ignoring the quark contribution upto NLO respectively as

$$G(x,t) = G(x,t_0)(t/t_0)^{B(x)} \qquad (12)$$

and

$$G(x,t) = G(x_0,t)t^{\{B(x)-B(x_0)\}}. \qquad (13)$$

here

$$B(x) = \frac{2}{\beta_0}f_3(x) + T_0\frac{2}{\beta_0}f_4(x),$$

$$f_3(x) = 6\times\left(\frac{11}{12} - \frac{N_f}{18} + \ln(1-x)\right) + 6\times\int_x^1 d\omega\left[\frac{(\omega^{\lambda+1}-1)}{1-\omega} + \left(\omega(1-\omega) + \frac{1-\omega}{\omega}\right)\omega^\lambda\right]$$

and

$$f_4(x) = \int_x^1 \left[\frac{52}{3}\omega^{\lambda-1}\right]d\omega.$$

**3 Results and discussions**

A new description of t and x-evolutions of gluon distribution function is presented through equations (9) and (10). Where we solved DGLAP evolution equation for gluon distribution function upto NLO considering Regge behaviour of distribution functions. In equations (12) and (13), we obtained the description of gluon distribution function ignoring the quark contribution in the DGLAP evolution equation for gluon distribution function upto NLO. The contribution of quark to gluon distribution functions theoretically should decrease for x→0, $Q^2$→∝. So we are interested to see the contribution of quark to gluon distribution functions in our region of discussion. We compare our result of t evolution of gluon distribution function upto NLO with GRV98NLO [13] global parametrization at $Q^2$=100 GeV$^2$ and the result of x-evolution with MRST2004 [14], GRV98LO [13] and GRV98NLO [13] global parametrizations at several medium to high-$Q^2$ range. Along with the NLO results we also presented our LO results [1]. We compare our results from the equations (9) and (10) for $K(x) = k$, $ax^b$ and $ce^{dx}$, where k, a, b, c and d are constants. In our work, we found the values of the gluon distribution function remains almost same for b<0.00001 and for d<0.00001. We have chosen b = d = 0.00001 for our calculation and the best fit graphs are observed by changing the values of k, a and c. As the value of λ should be close to 0.5 in quite a broad range of low-x [1, 8, 14], we have taken λ = 0.5 in our calculation.

In Figures 1(a) we plot $T(t)^2$ and $T_0T(t)$, where $T(t) = á_s(t)/2ð$ against $Q^2$ in the $Q^2$ range $0 \leq Q^2 \leq 200$ GeV$^2$ as required by our data used. Here we observe that for $T_0 = 0.05$, errors become minimum in the $Q^2$-range of our discussion $10 \leq Q^2 \leq 200$ GeV$^2$. From the graph it is clear that the difference between the values of $T(t)^2$ and $T_0T(t)$ in this range is negligible.

In figures 1(b) we compare our result of t-evolution of gluon distribution function upto NLO with GRV98NLO gluon distribution parameterization at $10^{-4}$. We compare our results for $K(x) = K$, $ax^b$ and $ce^{dx}$. At $x = 10^{-4}$, the best fit results are for k=a=c=0.4. The figures show good agreement of our result with GRV98NLO parameterization at low-x. In figures 2(a) to 2(d) we compare our result of x-evolution of gluon distribution function upto NLO each for $K(x) = K$, $ax^b$ and $ce^{dx}$ with GRV98NLO global parameterizations at $Q^2$ = 20, 40, 60, 100 GeV$^2$ respectively and best fit result corresponds to K=a=c= - 0.34 for $Q^2$ = 20GeV$^2$ and K=a=c= - 0.27 for $Q^2$ =40, 60 and 100 GeV$^2$. Figures 3(a) to 3(d) We have compared our result of x-evolution of gluon distribution function upto NLO with MRST2004 and GRV98LO global parameterizations each for $K(x) =ax^b$ and $ce^{dx}$ and with these results we have presented our LO results [1]. From both the results it is seen that the NLO results follows the parameterization graphs more closely than the LO results. Figures 4(a) to 4(d) show the sensitivity of the parameters λ, K=a=c, b and d respectively. Taking the best fit figures to the x-evolution of gluon distribution function upto NLO with GRV98NLO parameterization at $Q^2$ = 100 GeV$^2$, we have given the ranges of the parameters as 0.48≤λ≤0.52, -0.2≤K=a=c≤-0.34, 0.00001≤b≤0.01, and 0.00001≤d≤0.5.

## 4 Conclusions

In this paper we present an approximate analytical solution of the next-to-leading order DGLAP equation for the gluon structure function at low-x as a continuation of our earlier work at leading order [1]. We have considered the Regge behaviour of singlet structure function and gluon distribution function to solve DGLAP evolution equations. Here we find the t and x-evolutions of gluon distribution function upto NLO and compared with GRV98NLO global parameterization. We also compared our results for both LO and NLO with MRST2004 and GRV98LO global parameterizations and from the graphs it can be concluded that our results for NLO are in good agreement with MRST2004 and GRV98NLO and GRV98LO global parameterizations especially at low-x and high-$Q^2$ region. The x-evolution graphs obtained by solving DGLAP evolution equation upto LO and NLO can be compared and from the best fitted graphs it is clear that the gluon distribution function upto NLO shows significantly better fitting to the parameterizations than that of upto LO. So the higher order term upto NLO has appreciable contribution in the region of our discussion to the parton distribution function. The Regge behaviour of quark and gluon distribution functions is thus compatible with PQCD at that region. We were also interested to see the amount of contribution of quark to the gluon distribution function at different x and $Q^2$ but it has been observed that in our x-$Q^2$ region of discussion quark contributes appreciably to gluon distribution function even though we have taken the gluon distribution function upto NLO. But from the comparison of our results for x-evolution of gluon distribution function upto LO and NLO, it can be said that the contribution of quark to the gluon distribution function at fixed-$Q^2$ goes on decreasing with the inclusion of higher order corrections at low-x. So we cannot ignore the contribution of quark in our region of discussion. Considering Regge behaviour of distribution functions DGLAP equations thus become quite simple to solve.

**Figure captions**

**Fig. 1.** Fig. 1(a) is the variation of $T(t)^2$ and $T_0T(t)$ with $Q^2$ for $T_0=0.05$. Fig. 1(b) is best fit graphs of our result of t-evolution of gluon distribution function upto NLO for $\lambda = 0.5$, $T_0=0.05$ and $K(x) = K$, $ax^b$ and $ce^{dx}$ for the representative values of x presented with GRV98NLO parameterization $x=10^{-4}$. Data points at lowest-$Q^2$ values are taken as input to test the evolution equation (9). )

**Fig. 2.** Fig. 2(a)-2(d) are the best fit x-evolution graphs of our result with GRV98NLO parameterization for $\lambda = 0.5$, $T_0=0.05$, $K(x) = K$, $ax^b$ and $ce^{dx}$ and $Q^2 = 20, 40, 60$ and $100$ GeV$^2$ respectively. Data points at x = 0.1 are taken as input to test the evolution equations (10).

**Fig. 3.** Fig. 3(a) is the best fit x-evolution graph of our result with MRST2004 parameterization for $Q^2 =100$ GeV$^2$. Data points at x = 0.02 is taken as input to test the evolution equations (10). Fig. 3(b)-3(d) are the best fit x-evolution graphs of our result with GRV98LO parameterization for $Q^2 = 20, 40$ and $80$ GeV$^2$ respectively. Data points at x = 0.01 are taken as input to test the evolution equations (10). Here graphs are observed for $\lambda = 0.5$, $T_0=0.05$ and $K(x) = K$, $ax^b$ and $ce^{dx}$. Along with the NLO result we presented our LO results also.

**Fig. 4.** Fig 4(a) to 4(d) are the sensitivity of the parameters $\lambda$, K=a=c, b and d respectively at $Q^2 = 100$ GeV$^2$ with the best fit graph of our results with GRV98NLO parameterization.

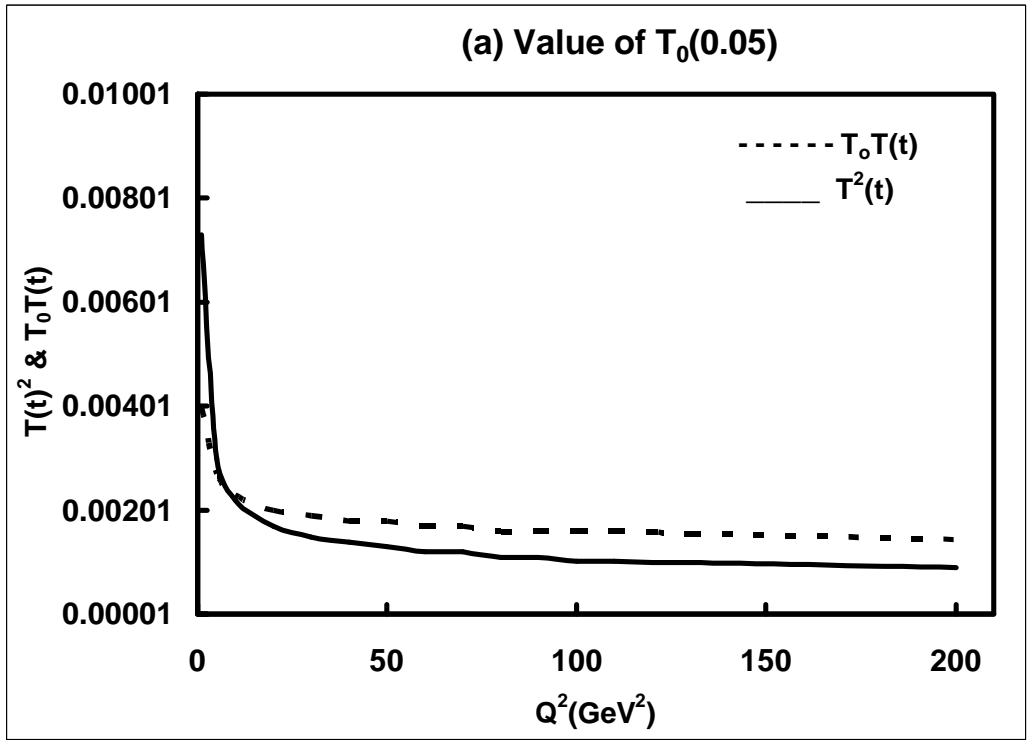

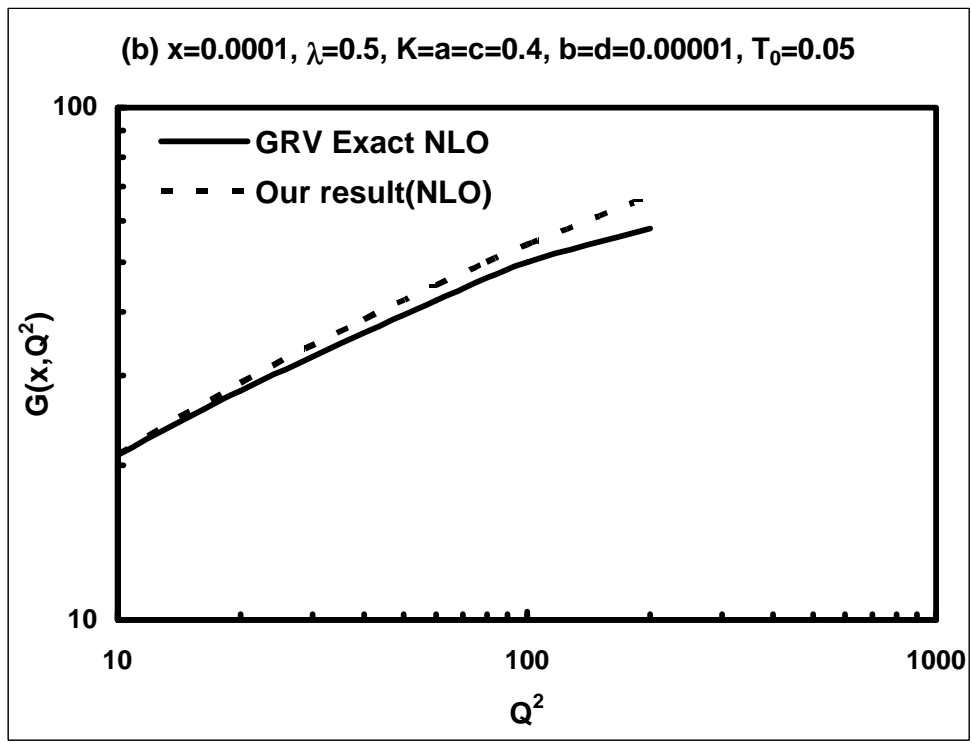

**Fig.1**

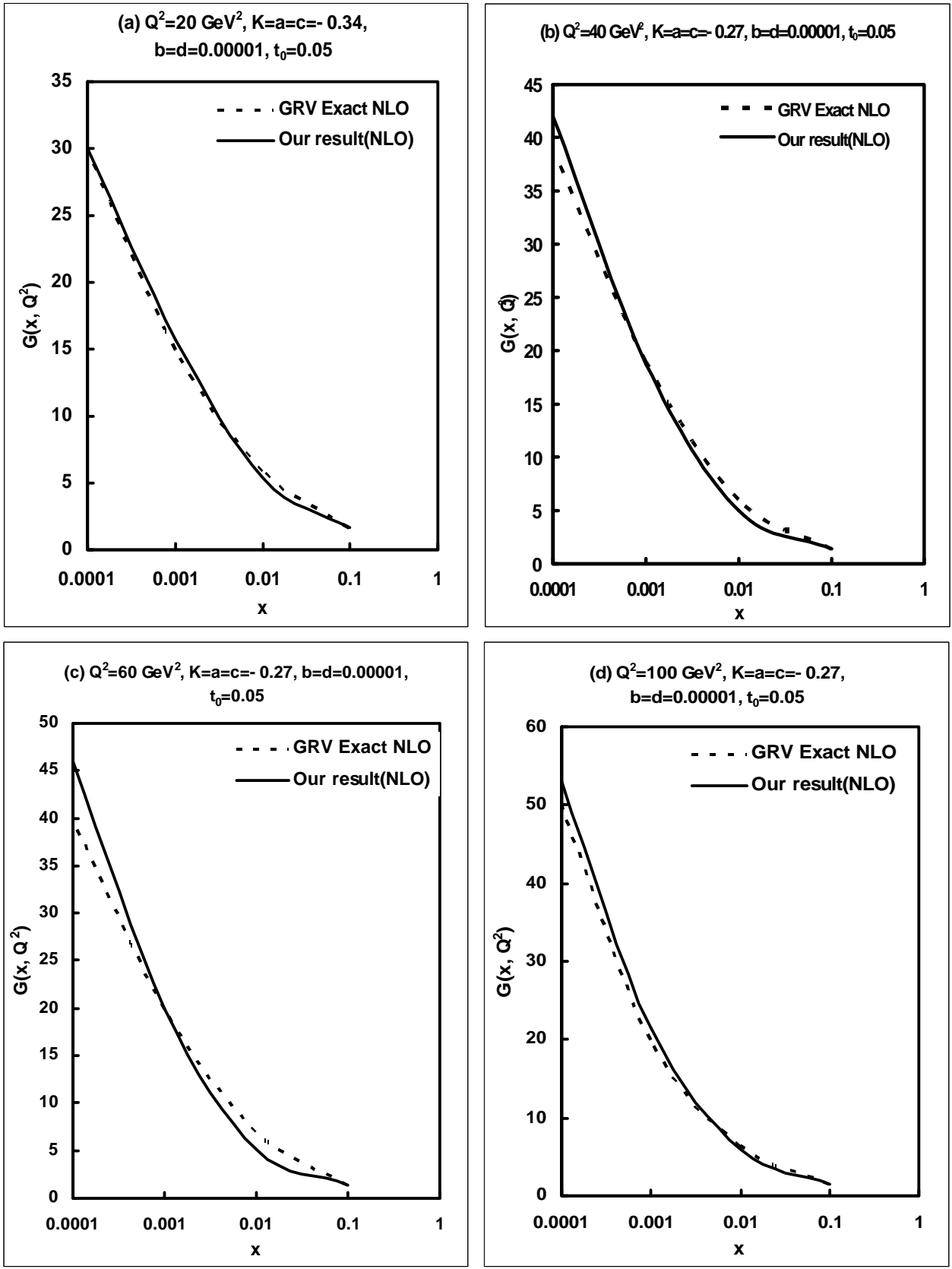

**Fig.2**

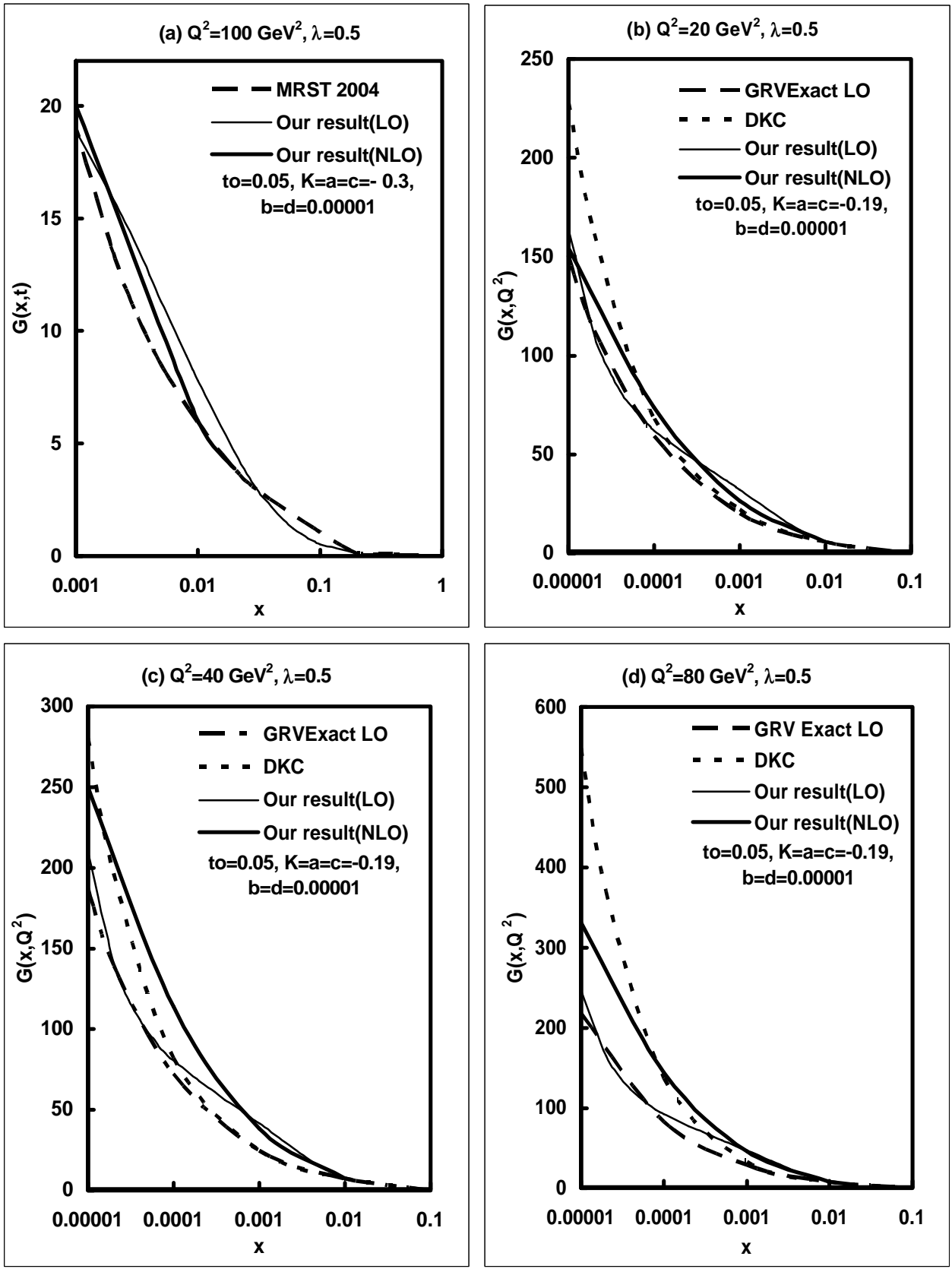

**Fig.3**

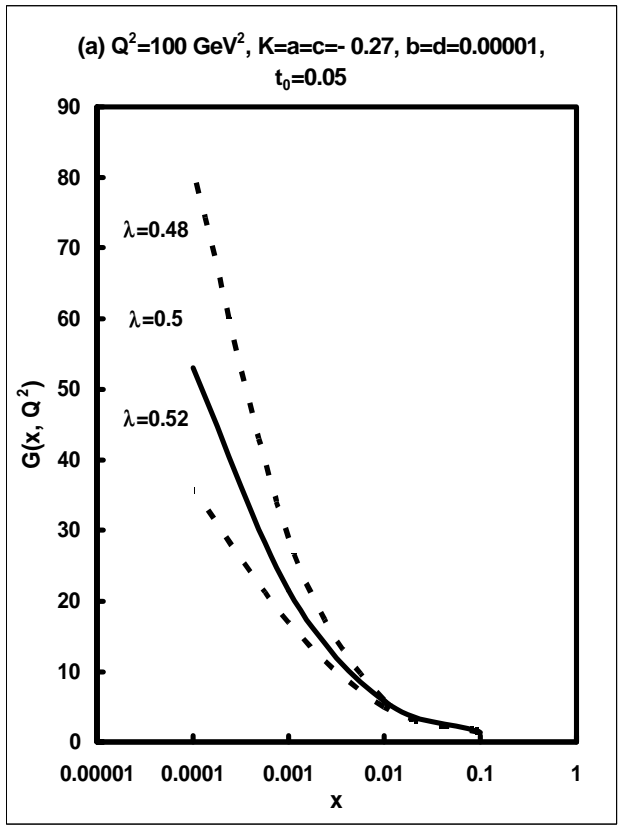
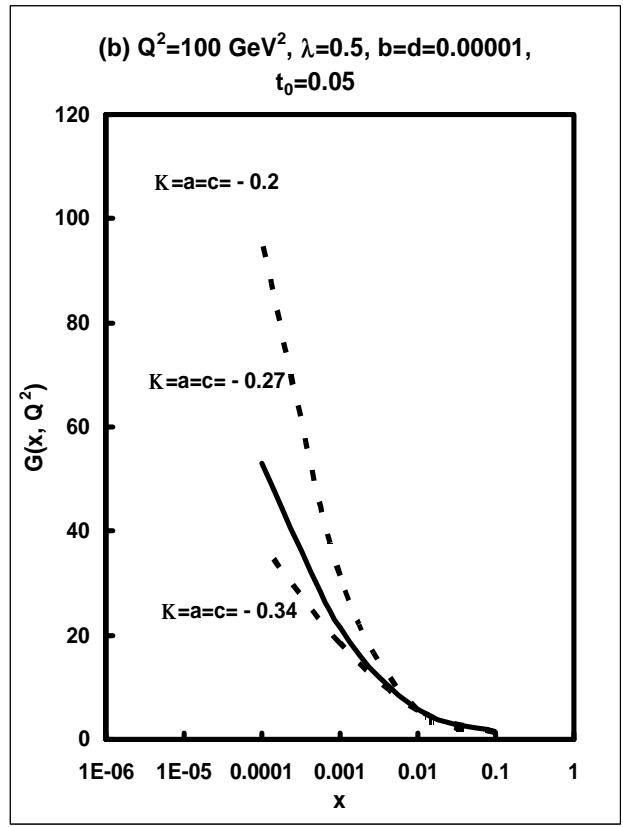
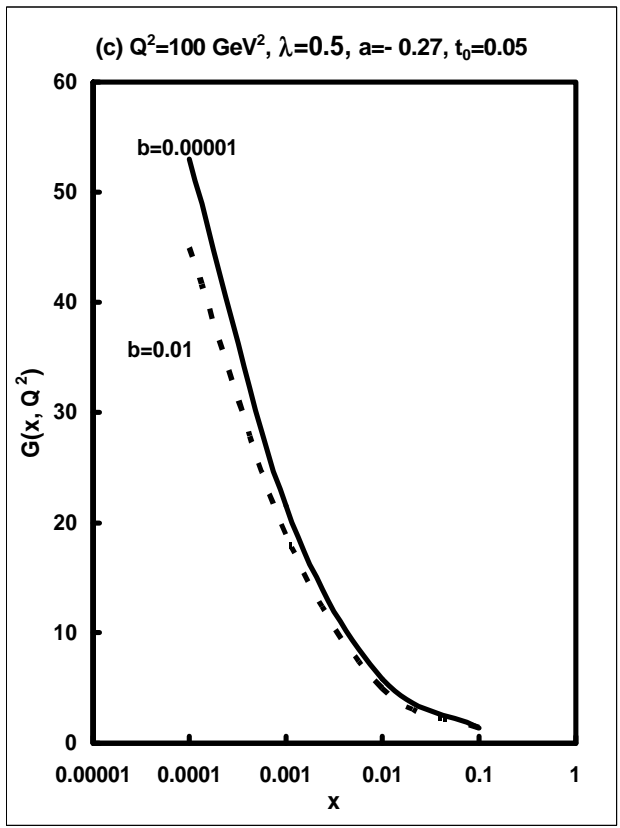
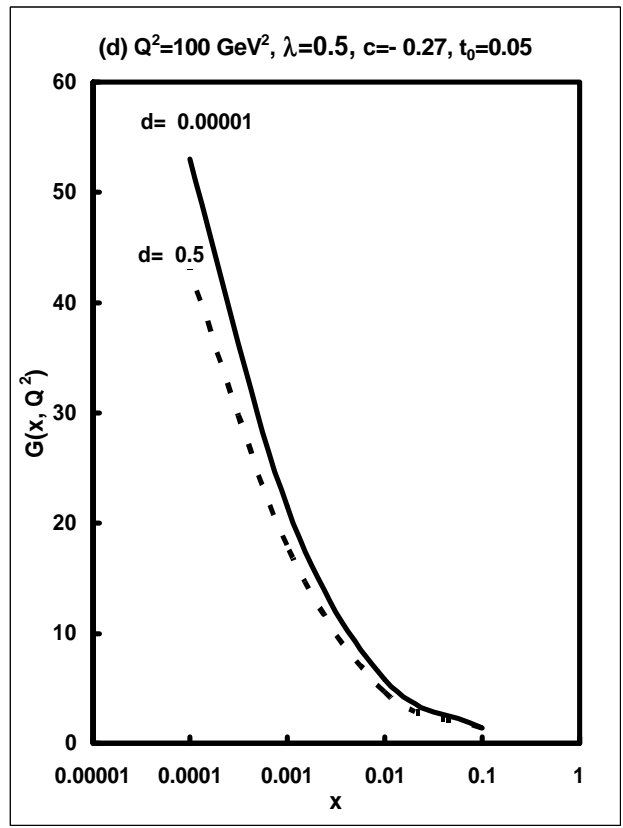

**Fig. 4**